\documentclass[aps,pre,showkeys,twocolumn]{revtex4-1}
\usepackage{color}
\usepackage{graphicx}

\begin{document}

\title{Fibril elongation mechanisms of HET-s
prion-forming domain: Topological evidence
for growth polarity}

\author{Marco Baiesi}

\author{Flavio Seno}

\author{Antonio Trovato}

\email[Author to whom correspondence should be addressed. \\ Phone
number: +390498277159. Fax number: +390498277102. \\ Electronic address: ]{trovato@pd.infn.it}

\affiliation{Department of Physics, Padua University - CNISM,
Padova Unit \\ Via Marzolo 8, I-35131 Padova, Italy}

\keywords{Protein aggregation; Dock and lock mechanism; Coarse-grained
models; Monte-Carlo simulations; Topology-based models; }

\begin{abstract}
The prion-forming C-terminal domain of the fungal prion HET-s forms
infectious amyloid fibrils at physiological pH. The
conformational switch from the non-prion soluble form to the prion
fibrillar form is believed to have a functional role, since HET-s in
its prion form participates in a recognition process of different
fungal strains. Based on the knowledge of the high-resolution
structure of HET-s(218-289) (the prion forming-domain) in its
fibrillar form, we here present a numerical simulation of the fibril
growth process which emphasizes the role of the topological properties
of the fibrillar structure. An accurate thermodynamic analysis of the
way an intervening HET-s chain is recruited to the tip of the growing
fibril suggests that elongation proceeds through a
dock and lock mechanism. First, the chain docks onto the fibril by
forming the longest $\beta$-strands. Then, the re-arrangement in the
fibrillar form of all the rest of molecule takes place.
Interestingly, we predict also that one side of the HET-s fibril is
more suitable for substaining its growth with respect to the
other. The resulting strong polarity of fibril growth is a consequence
of the complex topology of HET-s fibrillar structure, since the
central loop of the intervening chain plays a crucially different role
in favouring or not the attachment of the C-terminus tail to the
fibril, depending on the growth side.
\end{abstract}

\maketitle

\section{Introduction}

Prions are self-propagating, usually amyloid-like protein
aggregates\cite{Aguzzi2008} that are responsible for transmissible
diseases.  Examples of prion diseases are scrapie in
sheep\cite{Prusiner1982}, bovine spongiform
encephalopathy\cite{Wells87}(BSE) in cattle and new variant
Creutzfeldt-Jakob disease\cite{Will96} in man. In contrast to these
disease-related cases, several proteins displaying prion properties
are well known in yeast and fungi\cite{Serio2010}. The HET-s prion
protein produced by the filamentous fungus {\it Podospora anserina} is
thought to be involved in a specific function. The
switching on of its prion form triggers the programmed-cell-death
phenomenon named ``heterokaryon
incompatibility''\cite{Lange2008,Saupe2000} which can prevent
different forms of parasitism, by inducing the death
of the heterokaryon formed by cell fusion of different fungal
strains. The proteinase K-resistant core of the prion fibrils formed
by the C-terminal residues 218-289 (PFD:prion forming domain) is
unstructured in solution and forms infectious fibrils in
vitro\cite{Baulgerie2003}. Earlier work showed that HET-s PFD fibrils
consist of four $\beta$-strands forming two windings of a
$\beta$-solenoid\cite{Ritter2005} however without giving further
information about the details of the intra- and inter-molecular
$\beta$-sheet architecture. Recently, the structure of HET-s PFD has
been determined\cite{Wasmer2008} on the basis of NMR-derived intra and
intermolecular distance restraints\cite
{Castellani2002,Lange2005}. This is the only atomic-resolution
structural model of an infectious fibrillar state reported to date. On
the basis of 134 intramolecular and intermolecular experimental
distance restraints, HET-s PFD forms a left-handed $\beta$-solenoid,
with each molecule forming two helical windings, a compact hydrophobic
core, at least 23 hydrogen bonds, three salt bridges and two
asparagine ladders (see Fig. 1a and 1b). The model is supported by
electronic diffraction and scanning transmission electronic
microscopy\cite{Wasmer2008}.

Despite the tremendous economical and social relevance of diseases
related to prion infections and protein aggregation and the enormous
quantity of research devoted to their study, the mechanisms that rule
and trigger fibril formation and elongation remain
mostly elusive: the nature of protein aggregation, its limited
structural order, its insolubility in water and its involvement with
the cell membrane render its experimental study extremely
difficult. For these reasons relevant breakthroughs in understanding
the principles of amyloid formation and fibril
growth might come from numerical simulations.

Recent advances in hardware and methodology have allowed for realistic
atomic resolution molecular dynamics (MD) simulations with
physics-based potentials of small fibrils consisting of a few monomer
units that make possible to span time scales of hundreds of
nanoseconds \cite{Cecchini2006,Huet2006,Ma2006,
Tarus2006,Buchete2007,Baumketner2007, Brien2009} and even of $\mu s$
for smaller systems \cite{Reddy2009,Wei2010,Takeda2010}. These results
were mainly obtained by studying the amyloid $\beta$-peptide
(A$\beta$), related to the Alzheimer's disease\cite{Hardy2002}, or its
mutants and fragments. In this context, experimental
observations had first originated the proposal that incoming A$\beta$
monomers associate to the elongating fibril through a two stages
``dock and lock'' kinetic mechanism \cite{Esler2000}. In the first
stage, an unstructured monomer docks onto the fibril while maintaining
some degree of conformational freedom. In the second stage, the
monomer locks into the final fibril state. Such a mechanism was later
confirmed and refined, also in the context of A$\beta$-peptide
oligomerization, by other experimental and simulation studies
\cite{Cannon2004,Brien2009,Thir2010}. The dock and lock mechanism was
further employed, for the A$\beta$-peptide, to describe fibril
elongation within a thermodynamic context, by means of all-atom MD
simulations \cite{Takeda2009}. A continuous docking stage was observed
to occur over a wide temperature range without free energy barriers or
intermediates, whereas the locking stage at lower temperatures was
characterized by many competing free energy minima in a rugged
landscape.

In the case of the fibrillar HET-s PFD protein, it was possible with
MD atomistic simulation to probe the stability of the NMR structure on
a 10 ns timescale and to predict the behaviour of the salt bridge
network\cite{VanGusteren2009}. On the other hand, typical elongation
times for amyloid fibrils formed by HET-s PFD are of the order of
hours\cite{Sabate2009}, so that coarse grained approaches, in which
protein chains and amino-acid interactions are modeled in a simplified
way, are mandatory to investigate such longer time scales. Indeed,
despite the difficulty of finding reliable energy functions, these
approaches has been successfully used in outlining general aspects of
the full phase diagram of a generic aggregating polypeptide
system\cite{Hoang2006,Auer2009,Auer2010}, to emphasize the importance
of the contribution of hydrophobic interactions and hydrogen bonding
in the aggregation process of the A$\beta$ peptide
peptide\cite{Nguyen2006,Borroguero2005,Wei2004,Fawzi2005,Urbanc2004,
Favrin2004} and even to study the mechanisms of monomer addition for
the A$\beta$ peptide and some of its
mutants\cite{Lam2008,Urbanc2010,Rojas2010}.

In this paper, in order to describe the fibril
elongation mechanisms of a relatively long protein domain such as
HET-s PFD, we prefer to employ a still different coarse-grained
approach since in this case there is the unique advantage of knowing
from experiments the fibril structure. At a general level, our
strategy falls in the class of approaches used in protein folding that
builds on the importance of the native state topology in steering the
folding process \cite{Go76}. In its simplest example, the formation of
contacts is favoured only for pairs of amino acids that are found
interacting in the native state, but non-native sequence dependent
interactions could be introduced as well. Despite the semplicity of
this scheme, in the past few years, an increasing number of
experimental and theoretical studies have confirmed the utility of the
Go-like approach in the characterization of various aspect of protein
folding and binding
processes\cite{Micheletti99,Alm99,Baker2000,Clementi2000,Hubner2004,Levy2005,Lam2007,Sulk2009}.

The study of protein aggregation has been already tackled by using
Go-like models\cite{Ding2002,Jang2004}, but due to the absence of
experimental information on the fibrillar structure, hypothetical
aggregate conformations had to be introduced to build the Go-energy
function driving aggregation. Moreover, topologically based models,
with reduced effect of non-local interactions, correspond to funneled
energy landscapes and therefore their application should be limited to
situations in which proteins are evolutionarily designed to follow the
principle of minimal frustration\cite{Bryngelson87}, which results in
a faster search through the many possible alternatives. In general, a
funneled energy landscape is not expected in the case of
non-functional fibril formation. Therefore, the accurate knowledge of
its structure, the complex intra-chain topology, and the plausible
involvement in a functional process makes HET-s PFD a suitable and, at
the moment, unique candidate to exploit successfully Go-like models
for studying amyloid formation and fibril elongation
mechanisms.

Within this approach, implemented through a Monte-Carlo procedure
combined with replica exchange methods, we analyze the full
thermodynamic properties of the fibril elongation process, e.g. of the
association of a free chain to the already formed
fibrillar structure of HET-s PFD under different concentration
conditions. The behaviour of both energies and heat capacities shows
that the association process becomes more
cooperative for concentrations in the range ($\sim 10 \mu {\rm M}$) of
standard in-vitro experiment\cite{Greenwald2010}. A careful study of
the association process shows that fibril
elongation is triggered by the docking of the free chain onto the
fibril in a concentration dependent mechanism that
involves the formation of both inter- and intra-chain hydrogen bonds
stabilizing the longest $\beta$-strands, rapidly followed by the
assembly of the full domain. This behaviour is similar to the ``dock
and lock'' mechanism proposed for the amyloid A-$\beta$ fibril
formation \cite{Esler2000,Takeda2009}.

Another interesting aspect emerges clearly: elongation proceeds
differently according to which side of the fibril (see Fig 1a and 1b)
is used as the growing end. Elongation from one side is clearly
favoured with respect to the other side, implying a strong polarity in
the growth of HET-s PFD fibrils, which may even lead to unidirectional
elongation. Polarity in fibril growth is a feature already discussed
in the literature for other amyloid-forming proteins, in both
experiments\cite{Wilkinson2000,Weissman2002} and numerical
simulations\cite{Headgordon2007,Takeda2009}. Our data suggest that
growth polarity can be explained for HET-s PFD on the basis of the
complex topological properties of its fibrillar structure. A key role
is played by the behaviour of one long loop connecting two
$\beta$-strands in consecutive rungs of the fibrillar
structure. Depending on the elongation side, this loop may help or not
the winding of the C-terminal part of the attaching chain into the
fibrillar form. Since it is known that the prion loses its infectivity
upon partial deletion of that loop\cite{Ritter2005}, we argue that
this topological mechanism may be important for functional fibril
growth.
\section{Methods}

\subsection{Protein chain representation and energy function}

Chains A and B were selected from the NMR structure of HET-s PFD (PDB
code: 2rnm), where chain B stays on top of chain A along the fibril.
We keep one chain frozen whereas the other one is free. In the top
elongation mode, chain B is free and chain A is frozen, whereas in the
bottom elongation mode, chain A is free and chain B is frozen. The
free chain is allowed to move in a hemisphere of radius $L$ defined by
$x^2+y^2+z^2<L^2$, $0<z<L$ ($-L<z<0$) for top (bottom) elongation.
The frozen chain is placed with center-of-mass (i.e. average
C$^{\alpha}$) coordinates $x_{cm}=y_{cm}=0$ (i.e. on the hemisphere
axis) and $z_{cm}=+5.60$ {\AA} ($z_{cm}=-2.35$ {\AA}) for top (bottom)
elongation, rotated in such a way that the fibril axis is parallel to
the hemisphere axis.  The center-of-mass position along the hemisphere
axis is chosen in order to expose only one `sticky' end of the frozen
chain to the free chain, by placing the other end roughly on the
hemisphere base $z=0$. This allows a smaller computational effort, at
the expense of prohibiting conformations that we do not expect to
affect in a relevant way the binding of the free chain on the exposed
end of the full fibril, here represented by the frozen chain. The
portion of HET chain that we simulate goes from position 217 to
position 289, which includes 73 aminoacids (e.g. in the simulation we
include also the MET aminoacids at position 217 used to obtain the NMR
structure). We thus simulate a system with 146 aminoacids.

In order to perform extensive simulations, we use a coarse-grained
representation of the protein chain coupled with an energy function
based on the knowledge of the PDB fibril structure. In the spirit of
Go-like approaches widely used for globular proteins~\cite{Go76}, the
energy function is built in such a way to have its minimum for the PDB
structure. Inspired by Ref.~\cite{Hoang2004}, each aminoacid is
represented by an effective spherical atom located at the position of
the corresponding C$^{\alpha}$ atom. The virtual bond angle associated
with three consecutive $C^{\alpha}$ atoms is constrained between
$82^{\circ}$ and $148^{\circ}$.  Virtual bond lengths are kept
constant and equal to the native values from the PDB file. To
implement steric constraints we require that no two non-adjacent
$C_{\alpha}$ atoms are allowed to be at a distance closer than
$3.9$ {\AA}. We assign an energy $-1$ to each hydrogen bond that can be
formed between two residues that form it in the PDB fibril structure,
and we disregard any other attractive interaction (i.e. hydrogen bonds
cannot be formed by two residues that do not form it in the PDB fibril
structure and we do not consider any other type of pairwise
interaction except for the excluded volume constraints).  Only
$\beta$-sheet stabilizing hydrogen bonds can therefore be formed in
our simulation, and in order to identify them within a
$C^{\alpha}$-representation, we use the geometrical rules for
non-local hydrogen bonds introduced in Ref.~\cite{Hoang2004}. In order
to recognize the hydrogen bonds present in the PDB fibril structures
the lower bound on the scalar product between binormal and connecting
vectors was decreased to $0.88$ ($0.94$ in the original formulation,
see Table 1 and Fig. 1 in Ref.~\cite{Hoang2004} for the precise
listing of all hydrogen bond rules and the definition of binormal and
connecting vectors).

In this way, we find in the PDB fibril structure (see Fig. 2 and
Fig. 3) two long parallel $\beta$-strands ($\beta_1$ and $\beta_3$),
connected by 9 hydrogen bonds and four shorter strands: $\beta_{2a}$
parallel to $\beta_{4a}$ (linked together by 4 hydrogen bonds) and
$\beta_{2b}$ parallel to $\beta_{4b}$ (2 hydrogen bonds). Those
strands alternate within the fibrillar structure in pairs which form
hydrogen bonds within the same chain (intra-chain bonds) and pairs
which form hydrogen bonds between neighbouring chains (inter-chain
bonds). Each strand in the core of the fibrillar structure forms
intra-chain bonds on one side and inter-chain bonds on the other
side. In the ``top'' side of the fibril (see Fig. 2 and Fig. 3), the
exposed strands in the `sticky' end are $\beta_3$, $\beta_{4a}$
$\beta_{4b}$, whereas in the ``bottom'' side the exposed strands are
$\beta_1$, $\beta_{2a}$ $\beta_{2b}$.

Since we keep fixed one chain, the ground state has energy -30 (15
intra-chain hydrogen bonds plus 15 inter-chain hydrogen bonds - we are
not counting the intra-chain bonds of the frozen chain). Hence, energy
can range from $0$ (unbound chains) to $-30$ (fully bound chains in
the fibrillar state). In order to fix a realistic temperature scale,
the effective value of hydrogen bond energy in our Go-like energy
function was given the value $3.5\ {\rm Kcal} \cdot {\rm
mol}^{-1}$, so that the unit temperature of our simulations
corresponds to $1760$ K.

We have simulated the elongation of the fibril from both sides. For
each side we simulated chains confined in a hemisphere centred in the
origin and of radius $L_1=5\ {\rm nm}$, $L_2=10 \ {\rm nm}$, and
$L_3=30 \ {\rm nm}$ corresponding to concentrations $c_1=6.4 \ {\rm
mM}$, $c_2=0.8 \ {\rm mM}$ and $c_3= 30 \ \mu {\rm M}$. The latter
value is close to typical concentrations used in vitro aggregation
experiment~\cite{Sabate2009,Greenwald2010}.

\subsection{Monte-Carlo simulation}

Fibril elongation is simulated by means of a Monte Carlo procedure.
Multiple Markov processes~\cite{Whittington1996}, each replicating the
same system of a fixed chain and a free chain attaching to it
described above, are run in parallel at different temperatures, with
possible swaps of replicas, in order to sample efficiently the
conformational space from high ($T=210$ $^{\circ}$C) to low
temperatures ($T=10$ $^{\circ}$C). Within the same replica,
conformations are evolved using trial moves, which either pivot a part
of the chain from a randomly chosen residue to its end, or rotate a
chain portion between two residues along the direction joining
them. In the latter case the two residues are either chosen randomly
or chosen to be next-nearest neighbours along the chain. Trial moves
are accepted or rejected according to the Metropolis
test~\cite{Metropolis1953}. We use 20 different replicas, chosing
their temperatures to sample more accurately the transition region
($40$ $^{\circ}$C $<T<100$ $^{\circ}$C) and to provide reasonable
overlap of energy histograms between neighbouring pairs. Roughly
$2\cdot N$ Monte Carlo steps ($N=73$ is the number of residues of the
free chain) are performed independently for each
replica before one replica swap is attempted among a randomly chosen
neighbouring pair. Overall, roughly $10^7$ replica swaps are attempted
for each simulation and the acceptance rate of replica swaps is in all
cases above $60$\%. The convergence to the equilibrium regime is
assessed by looking at the evolution of system energy as a function of
Monte Carlo steps. In order to compute equilibrium thermodynamic
averages, the simulation portion corresponding to the initial $N_{eq}$
swaps is discarded from the collected data, with $N_{eq}$ ranging from
$1.5\cdot10^6$ to $5\cdot10^6$, depending on concentration. Data
from all temperatures are elaborated with the multiple-histogram
method~\cite{Wang2001}. In Fig. 4 some snapshots of the simulations
are represented.

\section{Results}

\subsection{Cooperativity of the elongation process}

To characterize the thermodynamic properties of the HET-s fibril
growth process we study the behaviour of the energy and of the heat
capacity of the system as a function of the temperature for three
different protein concentrations (see Fig. 5) and by considering
elongation both from the top and from the bottom side (see Methods).
The peaks in the heat capacity curve signal the occurrence of large
conformational rearrangements (that strictly speaking would become
transitions only in the thermodynamic limit of the number of system
component going to infinite) related to the process of the
deposition of the free HET-s PFD chain to the one
that is kept fixed to represent the sticky end of the already formed
fibril.

As expected, the first association temperature,
related to the high temperature heat capacity peak, decreases when
lowering the concentration. It reaches values close
to room temperature for the concentration $c_3=30 \ \mu {\rm M}$.
Interestingly, the cooperativity of the transition increases for lower
concentrations: this is signalled by a more sigmoidal behaviour of the
energy profiles, by the sharpening of heat capacity peaks and by the
almost complete merging in one single narrow peak of the otherwise
complex peak structure. Despite at a first glance the growth processes
from the two different sides look similar, one can notice that at low
concentration a residual peak remains at low temperature for the top
elongation case.

\subsection{Stability of hydrogen bond formation}

To elucidate further these behaviours and to understand the nature of
the conformational rearrangements related to heat capacity peaks we
computed the formation probability of hydrogen bonds for the different
strands as a function of temperature (shown in Fig. 6).
We define the stabilization temperature of a group of
hydrogen bonds to correspond to their average formation probability at
thermodynamic equilibrium being equal to 0.5. We compute stabilization
temperatures for 6 possible groups of hydrogen bonds, corresponding to
the different strand pairings shown as separate white/black bands in
Fig. 6. The resulting stabilization temperatures are summarized in
Table \ref{T1}.

We will use the term `fibrillar' for inter-chain hydrogen bonds. As an
example, strand $\beta_1$ of the mobile chain couples with strand
$\beta_3$ of the fixed chain, in the case of elongation from the top
side; whereas strand $\beta_3$ of the mobile chain couples with strand
$\beta_1$ of the fixed chain in the case of elongation from the bottom
side, and similarly for other strand pairs (see
Fig. 6 caption for a detailed explanation of how intra-chain and
fibrillar hydrogen bonds are represented in the figure).

For all concentrations and for both elongation sides, the first
hydrogen bonds which become stable are the inter-chain ones formed
between the long strands $\beta_1$ and $\beta_3$, at a temperature
$T_{f13}$ varying from $115$ $^{\circ}$C for top elongation at
concentration $c_1$, to $65$ $^{\circ}$C for bottom elongation at
$c_3$ (see Table \ref{T1}). The stabilization temperature $T_{f13}$ of
fibrillar hydrogen bonds between the long strands is higher for higher
concentrations and, at the same concentration, for the case of top
side elongation.

This first stabilization process is followed at a lower temperature
$T_{i13}$ (varying from $85$ $^{\circ}$C for top elongation at
concentration $c_1$, to $63$ $^{\circ}$C for bottom elongation at
$c_3$, see Table \ref{T1}) by the stabilization of intra-chain
hydrogen bonds formed between the two long strands, $\beta_1$ and
$\beta_3$, of the mobile chain. This second stabilization temperature
is again higher for the top elongation case, at the same
concentration, and increases with concentration.

With a further decrease of the temperature the fibrillar hydrogen
bonds formed between the short strands ($\beta_{2a}$ with $\beta_{4a}$
and $\beta_{2b}$ with $\beta_{4b}$) become stable. The stabilization
temperature $T_{f24a}$ for inter-chain strand pair $\beta_{2a}$ and
$\beta_{4a}$ does not depend on concentration but only on the
elongation side, being higher, $T_{f24a}\simeq63$ $^{\circ}$C, for top
elongation with respect to $T_{f24a}\simeq50$ $^{\circ}$C for bottom
elongation. The stabilization temperature $T_{f24b}$ for inter-chain
strand pair $\beta_{2b}$ and $\beta_{4b}$ is roughly the same,
$T_{f24b}\simeq42\div43$ $^{\circ}$C, in all cases.

The last step involves the stabilization of the intra-chain hydrogen
bonds between the short strands and do not depend as
well on concentration. At this stage the more
significant difference between the two elongation sides emerges.
For top side elongation, the two short strand pairs are stabilized
roughly together at a temperature much lower with
respect to the previous step ($T_{i24a}\simeq10$
$^{\circ}$C corresponding to the small peak in the heat capacity curve
and $T_{i24b}\simeq-2$ $^{\circ}$C). For bottom side elongation, the
stabilization of the two short intra-chain strand pairs takes place at
quite different temperatures: hydrogen bonds between strands
$\beta_{2a}$ and $\beta_{4a}$ are stabilized at a
even slightly higher temperature than their inter-chain counterpart
($T_{i24a}\simeq44$ $^{\circ}$C), whereas the intra-chain strand pair
$\beta_{2b}$ and $\beta_{4b}$ is stabilized at a much lower
temperature ($T_{i24b}\simeq6$ $^{\circ}$C).

\subsection{'Dock and lock' mechanism}

Upon assuming that the order in hydrogen bond stabilization mirrors a
similar order in the kinetics of the elongation process, we can
extract a general message from these results by stating that the
attaching of a mobile chain onto the elongating fibril is triggered by
the formation of the inter-chain hydrogen bonds of the longest strand
($\beta_1$ or $\beta_3$) followed by a first partial folding of the
chain through the formation of the long intra-chain strand pair
(between $\beta_1$ and $\beta_3$) and by the successive formation of
all other inter- and intra-chain hydrogen bonds. The mechanism of
addition of a soluble unstructured monomer to a preformed ordered
amyloid fibril is a complex process: the deposition involves an
association of the unstructured monomer to the fibril surface
(docking) followed by a conformational rearrangement leading to the
incorporation onto the underlying fibril lattice (locking)
\cite{Esler2000,Cannon2004,Brien2009,Thir2010}.

We identify the docking stage with the formation of both inter- and
intra-chain hydrogen bonds between the long strands $\beta_1$ and
$\beta_3$, as in both cases the stabilization temperatures $T_{f13}$
and $T_{i13}$ depends on concentration (see Table \ref{T1}).
One indeed expects that in a denser regime it is
easier for the mobile chain to dock onto the fibril end, while the
locking into the $\beta$-helix structure necessary for fibril
propagation should not be affected by concentration
changes. Therefore, the locking stage involves the formation of both
inter- and intra-chain hydrogen bonds between the remaining shorter
strands, since we observe that their stabilization
temperatures do not depend on concentration. The dependence of the
intra-chain hydrogen bond stabilization temperature $T_{i13}$ upon
concentration is non-trivial and is triggered by the strong
concentration dependence of the formation probability of the fibrillar
hydrogen bonds between the long strands.  The higher the
concentration, the more probable the fibrillar bonds to be
formed. Consequently, the more easily the intra-chain bonds are
stabilized.

From the data shown in Fig. 6 and in Table \ref{T1}, another general
feature can be picked out: the temperature range in which
the docking stage and thus the full elongation
process take place decreases at lower concentrations. This is
consistent with the increment of cooperativity as it appears from
thermodynamic quantities (e.g. sharpness of heat capacity peaks in
Fig. 5) when concentration diminishes. The most cooperative behavior,
as shown by the presence of a single sharp peak in the heat capacity
curve, is obtained at concentration $c_3$ in the case of bottom side
elongation, whereas a second peak is clearly visible at low
temperature for top side elongation at the same concentration. The
above analysis of Fig. 6 data reveals the crucial role of the
formation of intra-chain hydrogen bonds between strands $\beta_{2a}$
and $\beta_{4a}$ in this respect. The stabilization temperature
$T_{i24a}$ of these hydrogen bonds constitutes the
most relevant difference between the two elongation
modes in the first place. Moreover, it does correspond closely in both
cases to significant features in the heat capacity curve, namely the
small peak at $T_{i24a}\simeq10 \ ^{\circ}$C for top elongation and
the small shoulder in the main peak at $T_{i24a}\simeq44\ ^{\circ}$C
for bottom elongation.

\subsection{Competition between intra-chain and inter-chain hydrogen bonds}
\label{entr}

In order to gain a further understanding of the role played by
intra-chain $\beta_{2a}$-$\beta_{4a}$ hydrogen bonds we computed for
different temperatures the equilibrium occupation probabilities of
macroscopic conformational states, which are defined according to the
number of formed intra- or inter-chain hydrogen bonds. In Fig. 7 and
Fig. 8 the results are shown for concentration $c_3$ and for the two
different growth modes. Occupation probabilities are shown in
logarithmic scale, so that the resulting data could be interpreted as
(the opposite of) effective free energies or mean force potentials. At
high temperature, the bottom left corner is mostly populated,
corresponding to conformations with very few or none intra- and
inter-chain hydrogen bonds formed, typical of a mobile chain not yet
attached to the fibril end. On the other hand, at very low
temperature, the opposite top right corner is populated, describing
structures with almost all the hydrogen bonds formed that correspond
to mobile chains found already completely associated to the fibril end
with a significant probability.

Consistently with the previous analysis, based on Fig. 6 data, the
elongation process is complex, taking place in several stages.
Macrostates with only intra-chain hydrogen bonds are found to be
populated to some extent at high temperatures, hinting to the
possibility of a pre-structuring of the mobile chain before docking to
the fibril tip, yet the pathway more significantly visited involves
the population of first the 9 fibrillar hydrogen bonds between strand
$\beta_1$ and $\beta_3$ and then the analogous intra-chain bonds (see
$T=70\ ^{\circ}$C in Fig. 7 and Fig. 8).

After this first stage, that we already identified
with docking, two different scenarios emerge again clearly,
depending on the growth mode. The overall process is visibly more
cooperative for bottom side elongation (Fig. 8) with respect to top
side elongation (Fig. 7), since the spreading of significantly visited
macrostates is restricted to a narrower temperature range in the
former case. Moreover, around room temperature ($T=20-30\ ^{\circ}$C),
the most populated state for bottom side elongation has 15 fibrillar
and 13 intra-chain hydrogen bonds formed corresponding to a chain
almost completely attached to the fibril end (only strand $\beta_{2b}$
is left wiggling a bit). Instead, the most populated state for top
side elongation has all 15 inter- but only 9 intra-chain hydrogen
bonds locked into the fibrillar conformation, signalling again that
the stability of intra-chain $\beta_{2a}$-$\beta_{4a}$ hydrogen bonds
is strongly weakened with respect to bottom side elongation.

In order to appreciate more easily the variations
with temperature in the population of the different macrostates, we
computed unidimensional free energy profiles as a function of the
number of either intra- or inter-chain hydrogen bonds. The results are
shown in Fig. 9 for concentration $c_3$ and for the two different
growth modes.

Fig. 9 pictures confirm the multistage nature of the
association process with different macrostates that become the global
free energy minimum at different temperatures. The free energy
profiles as a function of the number of fibrillar hydrogen bonds are
similar in both elongation modes. The main difference is the value of
the temperature below which the free state (none fibrillar hydrogen
bond is formed) is not the free energy minimum anymore: $70\
^{\circ}$C for top elongation and $60\ ^{\circ}$C for bottom
elongation, consistently with Table \ref{T1}.

The free energy of the free state and the free energy
barrier that separates it from the competing minimum with $9$
fibrillar bonds (the ones formed between the $\beta$-strands $\beta_1$
and $\beta_3$ that are the first to be stabilized in the association
process) do not basically depend on temperature. This is a signature
of their entropic origin (free energies in fig. 9 are plotted in $RT$
units), as they are both dominated by the roto-translational entropy
of the free chain.

On the other hand, the free energy profiles as a
function of the number of intra-chain hydrogen bonds display a
relevant difference between the two elongation modes, consistently
with previous observations. For top elongation, the free energy
minimum at $T=20\ ^{\circ}$C is the macrostate with only $9$
intra-chain hydrogen bonds (i.e. the $\beta$-strands $\beta_{2a}$ and
$\beta_{4a}$ are not yet paired), whereas for bottom elongation the
free energy minimum at $T=20\ ^{\circ}$C is the macrostate with $13$
intra-chain hydrogen bonds (i.e. the $\beta$-strands $\beta_{2a}$ and
$\beta_{4a}$ are already paired).

Interestingly, the above difference is due to the macrostate with $9$
intra-chain hydrogen bonds being entropically stabilized for top
elongation with respect to bottom elongation.  Indeed, the free
energies for $10\div15$ intra-chain hydrogen bonds do not differ in
the two elongation modes. Moreover, the free energy difference between
the two elongation modes for $9$ intra-chain hydrogen bonds has to be
entropic, since the energy of that macrostate is the same in both
cases, being given by the $9$ intra-chain plus the $15$ inter-chain
hydrogen bonds (the latter is the global free energy minimum at $T=20\
^{\circ}$C for both elongation modes).

\subsection{Topological origin of fibril growth polarity}

Our analysis clearly established that fibril growth exhibits a deeply
different thermodynamic behaviour depending on the side from which
elongation proceeds: at room temperature only bottom side elongation
is thermodynamically stable. What is the physical reason for the
existence of such a strong growth polarity?  Since our simulation
study is based only on the knowledge of the fibril structure and not
on the specificity of the amino-acid sequence, we can expect that
growth polarity is a consequence of the topological properties of the
structure.

There is indeed a deep topological difference between the deposition
mechanisms of HET-s PFD on the two different sides of the fibril. In
the first docking stage, common to both elongation modes, the
formation of both inter- and intra-chain hydrogen bonds between
$\beta_1$ and $\beta_3$ implies the positioning of the latter strands
into a conformation already compatible with the final fibrillar
structure. The remaining strands, then, yet to be positioned, acquire
distinct topological roles, since $\beta_{2a}$ and $\beta_{2b}$ are in
a loop between the two chain portions already pinned down in the
fibrillar structure, whereas $\beta_{4a}$ and $\beta_{4b}$ are in the
C-terminal tail of the chain. One can then predict that, for entropic
reasons, the former pair can be accomodated into the final
fibrillar structure more easily than the latter pair. Nevertheless,
different elongation modes may change this hierarchy.

When elongation proceeds from the top side, the attaching chain wraps
up onto the fibril tip starting with its N-terminal part (see
Fig. 10). The `loop' strands $\beta_{2a}$ and $\beta_{2b}$ (fibrillar)
need to be positioned before the `tail' ones $\beta_{4a}$ and
$\beta_{4b}$ (intra-chain), consistently with the topological order
suggested above. Indeed we do observe this hierarchy for top side
elongation; even at low temperature intra-chain hydrogen bonds between
the short strand pairs are not yet stable.

When elongation proceeds from the bottom side, the attaching chain
wraps up onto the fibril tip starting with the C-terminal part (see
Fig. 10). The `tail' strands $\beta_{4a}$ and $\beta_{4b}$ (fibrillar)
need now to be positioned before the `loop' ones $\beta_{2a}$ and
$\beta_{2b}$ (intra-chain). Elongation order takes over the
topological order so that the difficult positioning of `tail' strands
is assisted by the easier positioning of `loop' strands and both are
stabilized at room temperature (with the partial exception of the
shortest strand $\beta_{2b}$).

\section{Discussion}

In this work we have used Monte-Carlo simulations of a coarse-grained
representation of HET-s PFD domain in order to get information about
the elongation of the corresponding amyloid fibril by attaching of a
mobile chain to a pre-fixed fibrillar structure. Our approach, based
on the knowledge of the fibrillar structure, relies on the currently
well accepted assumption that protein topology plays a pivotal factor
in determining unimolecular folding and protein assembly. At variance
with other Go-like studies \cite{Ding2002,Jang2004}, based on a
hypothetical structure, the reliability of our study is justified by
the knowledge of a high-resolution NMR structure of a plausibly
functional amyloid fibril.

There are two main results of our thermodynamic study. First, we
observe that fibril elongation is driven by the formation of
inter-chain and intra-chain hydrogen bonds between the long strands
$\beta_1$ and $\beta_3$, followed by the positioning of the rest of
the attaching chain onto the growing fibril. This mechanism is known
as {\em dock and lock mechanism}
\cite{Esler2000,Brien2009}. We identify the docking
stage as that part of the association process whose onset temperature
displays a concentration dependance. A similar feature, i.e. the
docking temperature range varies with concentration, was previously
found in a thermodynamic study of A$\beta$-peptide fibril growth
\cite{Takeda2009}. Within our Go-like approach, we find a complex
multistage association process, where both the docking and the locking
stages proceed in several steps in a free energy landscape
characterized by several different minima separated by barriers. In
the case of A$\beta$-peptide fibril elongation, it was found on the
contrary that docking is continuous and proceeds without the presence
of intermediate or free energy barriers \cite{Takeda2009}. Our finding
of a multistage docking behaviour might be due to the non-trivial
intra-chain topology of HET-s PFD monomer, lacking in the
A$\beta$-peptide case. Alternatively, it might be an artifact caused
by our neglecting of non-native interactions.

Secondly, we predict that one side of the structure is more suitable to substain
the growth of the fibril. 

The predicted fibril growth polarity can be rationalized by analyzing
elongation topological properties, which turn out to be intrinsically
different from the two fibril sides. The argument is based on the
complex tertiary structure of the monomeric unit within HET-s fibril,
resulting into alternating intra- and inter-chain pairs of
$\beta$-strands. After the first hydrogen bonds have been formed in
the initial docking onto the growing fibril, the portion of the
attaching chain which is going to acquire $\beta$-structure in the
following locking stage, is partitioned into a `loop' and a `tail'
part (see Fig. 10). As a consequence, the entropy loss of the two
parts upon locking is different, implying a `topological'
hierarchy. The latter may or may not be affected by the `winding'
hierarchy dictated by the choice of the growth side, thus resulting in
a strong growth polarity. Bottom side elongation is more sustainable
because the `loop' part may assist the `tail' part to
lock. The entropic origin of the growth polarity
observed in our results is further confirmed by the free energy
profiles shown in Fig. 9. As discussed in section \ref{entr}, the
macrostate populated after docking and before locking (with $15$
inter-chain and $9$ intra-chain hydrogen bonds) is entropically
stabilized in the top elongation mode with respect to bottom
elongation.

Being based on a topological argument, we believe our prediction to be
robust against both variations of the details in the implementation of
our Go-like approach (changing the coarse-graining of protein chain
representation, employing in the energy function general pairwise
contacts and not only hydrogen bonds, using different hydrogen bond
rules) and relaxation of other simplifying assumptions that we made,
namely the fibril tip is represented by just one chain and kept
completely frozen. This latter point is further
motivated by the experimental observation \cite{Sabate2009} that the
HET-s fibril accomodates incoming prion monomers without a substantial
disorganization of its structure. This behaviour is quite different
from those of A$\beta$ fibrils where it was experimentally observed
\cite{Kusumoto1998} an entropy gain in the elongation reaction which
was related to an unfolding of the organized fibril end to accomodate
the addition of the incoming monomers.

We thus suggest our prediction may be accessible
to experimental validation, for instance using the
single fiber growth assay employed in \cite{Weissman2002} for SUP35
yeast prion, based on two variants of the prion domain that can be
differentially labelled and distinguished by atomic force
microscopy.

The same topological argument cannot be used for simpler amyloid
structure such as the solid state NMR-model suggested for the
Alzheimer's $A\beta$-peptide~\cite{Petkova2002}, in the absence of
intra-chain hydrogen bonds. In fact, experimental evidence shows
bidirectional fibril growth with no clear signs of growth
polarity\cite{Wilkinson2000}. Interestingly, the presence of
asymmetrical topological properties of fibril ends was indeed
suggested for the $A\beta$-peptide~\cite{Headgordon2007,Takeda2009},
depending on the different possible quaternary arrangements of the
fibril~\cite{Tycko2006}. On the other hand, a clear evidence of growth
polarity was demonstrated for SUP35 fibrils~\cite{Weissman2002}, a
yeast prion protein believed to have a functional role, similarly to
HET-s prion. There is no high-resolution information on the structure
of SUP35 fibrils and different conflicting structural models have been
recently proposed~\cite{Lindquist2005,Shewmaker2006}, yet some of the
available data suggests the existence of a complex intra-chain
structure~\cite{Lindquist2009}, which could justify the observed
growth polarity within a framework similar to the one we propose here
for HET-s.

A peculiar feature of HET-s PFD is the presence of a long loop in the
fibrillar structure that connects strands $\beta_{2b}$ and $\beta_3$
(aa 246-260), which then contributes to increasing the length of the
chain portion partitioned as `loop' in the topological argument
discussed above (in blue in Fig. 10). An interesting question is
whether increasing such length favours or disfavours growth polarity
according to the mechanism suggested here. One could argue that a
longer loop may further decrease the fluctuations available to the
`tail' part, thus enhancing its assisted locking for bottom side
elongation. Indeed, it is known that a large deletion in such loop
(248-256) led to loss of HET-s function and
infectivity\cite{Ritter2005}. Even though no evidence
is known about the possible biological relevance of fibril growth
polarity in fungal and yeast prions, it is tempting to speculate
that the role of the loop might be to enhance growth polarity as a way
to control the elongation process more thoroughly in the context of
the propagation of a functional prion. Numerical simulations performed
within the same Go-like approach, based on a model structure to be
built on the basis of HET-s PFD but with a shorter loop, will shed
further light on this hypothesis.

Our coarse-grained approach turned out to be effective in studying
structural rearrangements which could not be tackled using more
detailed protein chain representations. On the other hand, we are
neglecting factors which may play an important role in the elongation
process, such as the possible presence of oligomeric conformers of
HET-s PFD in the non-fibrillar soluble state~\cite{Sabate2009}. In
order to capture similar effects, one needs to increase the accuracy
of the energy function by introducing amino-acid specificity. This can
be done by using coarse-grained potentials which take into account the
different ability of each pair of amino-acids in forming hydrogen
bonds in $\beta$-strands~\cite{Trovato2006,Trovato2007}, coupled with
similar potentials describing residue pairwise
interactions~\cite{Miyazawa1996} or local conformational
biases~\cite{Shortle2005}. Such potentials might be used to modulate
the Go-like energy function presented in this work.
The use of a more realistic description of the
protein chain involving side chain atoms may also cause the amino-acid
sequence to affect the elongation process in a side-depending manner
by imposing chiral stereochemical constraints. The latter are not
present in our C$^{\alpha}$-based representation, thus reinforcing the
topological origin of growth polarity in our results.

Finally, the approach presented here could be also used to study the
nucleation process of HET-s PFD, and further extended to study the
aggregation of the full HET-s protein. The HET-s N-terminal domain in
the non-fibrillar soluble form is structured into a globular protein
fold whose high-resolution structure has been very recently
released\cite{Greenwald2010} (PDB code 2wvn.pdb). Such structure is
known to lose partially its order upon ordering and aggregation of
HET-s PFD into the unsoluble fibrillar form~\cite{Wasmer2009}. A Go-like
approach would be especially suited to study the resulting competition
between the ordered structures of the two domains in the two different
forms~\cite{Zamparo2010}.

\begin{acknowledgments}
We thank F. Chiti and R. Riek for enlightening discussions. We
acknowledge financial support from University of Padua through
Progetto di Ateneo n. CPDA083702.
\end{acknowledgments}

\newpage

\section*{Table Legends}

\begin{table}[!hb]
\begin{tabular}{|c|c|c|c|c|c|c|}
\hline & $T_{f13} (^{\circ}{\rm C})$ & $T_{i13} (^{\circ}{\rm C})$ &
$T_{f24a} (^{\circ}{\rm C})$ & $T_{f24b} (^{\circ}{\rm C})$ &
$T_{i24a} (^{\circ}{\rm C})$ & $T_{i24b} (^{\circ}{\rm C})$ \\ \hline
\multicolumn{7}{|c|}{top side elongation} \\ \hline $c_1$ & 115 & 85 &
64 & 43 & 12 & -1 \\ $c_2$ & 88 & 71 & 63 & 44 & 11 & -2 \\ $c_3$ & 75
& 70 & 62 & 43 & 8 & -3 \\ \hline \multicolumn{7}{|c|}{bottom side
elongation} \\ \hline $c_1$ & 108 & 80 & 50 & 42 & 45 & 6 \\ $c_2$ &
85 & 68 & 47 & 40 & 43 & 7 \\ $c_3$ & 65 & 63 & 50 & 41 & 44 & 6 \\
\hline
\end{tabular}
\caption{\label{T1} Stabilization temperatures of
hydrogen bond groups for the different concentrations and elongation
modes used in our simulations. Hydrogen bonds stabilizing a given
strand pair are grouped into the same set, resulting into 6 different
sets (see Fig. 6 caption for a detailed explanation). Stabilization
temperatures for different sets are labeled in such a way that, e.g.,
$T_{f13}$ ($T_{i24a}$) refers to fibrillar (intra-chain) hydrogen
bonds formed between strands $\beta_1$ and $\beta_3$ ($\beta_{2a}$ and
$\beta_{4a}$).}
\end{table}

\begin{figure*}[!ht]
\includegraphics[]{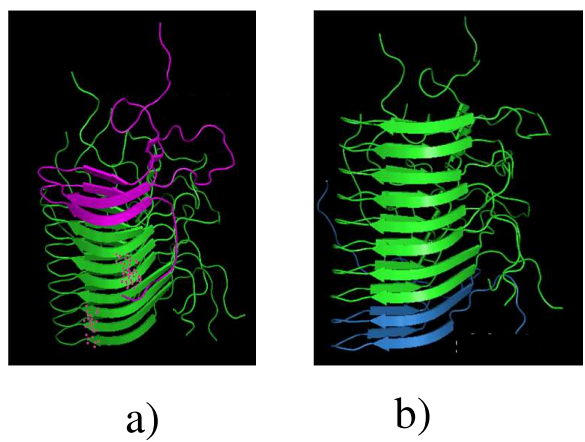}
\caption{\label{Fig1} {\bf High resolution experimental structure of
HET-s fibril.}  Side view of the five molecules present in the
fibrillar structure of HET-s PFD calculated with NMR restraints (PDB
code:2RNM). In a) the monomer on the top of the fibril (chain E) is
drawn in violet, whereas in b) the bottom monomer (chain A) is drawn
in blue}
\end{figure*}

\begin{figure*}[!ht]
\includegraphics[]{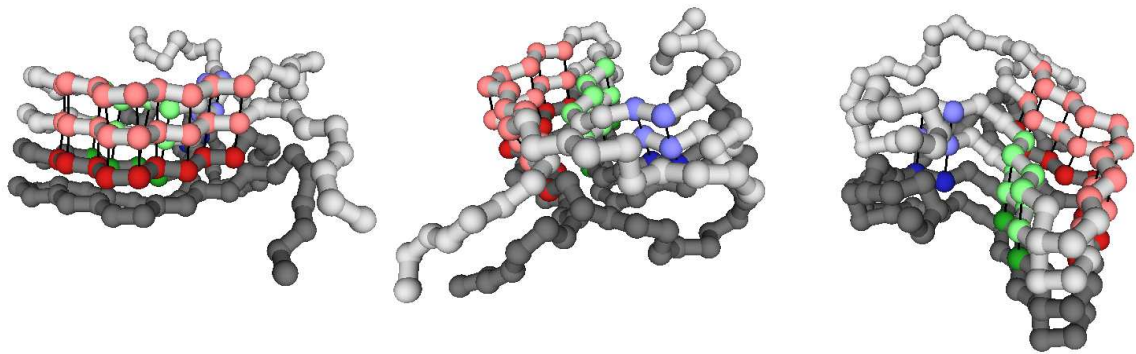}
\caption{\label{Fig2} {\bf Coarse-grained represention of HET-s
molecules (top side elongation).} Different views of the HET-s PFD
structure used as the minimum energy structure in our
simulations. Chains A and B from the PDB fibril structure are
kept. During the simulation of top side elongation, the light chain is
mobile while the dark one is frozen and represents the fibril end to
which the mobile chain is attaching.  The three types of
$\beta$-stranded regions are depicted in red ($\beta_1$ / $\beta_3$),
green ($\beta_{2a}$ / $\beta_{4a}$), and blue ($\beta_{2b}$ /
$\beta_{4b}$). The hydrogen bonds used in the Go-like energy functions
are represented by dark lines joining the interacting C$^{\alpha}$
atoms.}
\end{figure*}

\begin{figure*}[!ht]
\includegraphics[]{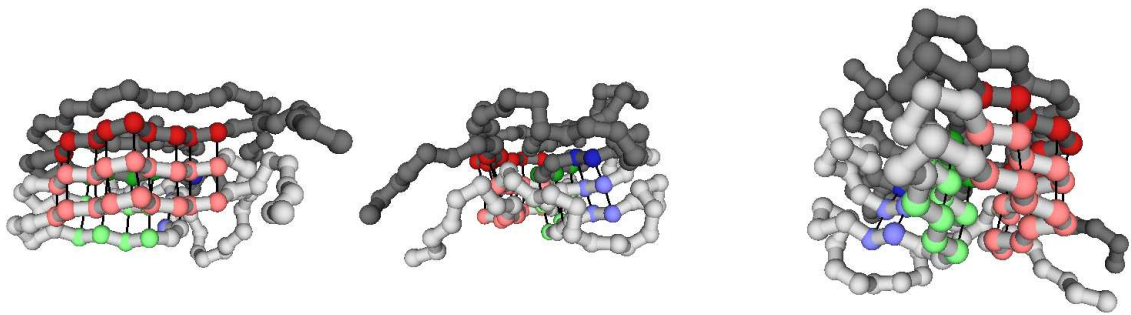}
\caption{\label{Fig3} {\bf Coarse-grained represention of HET-s molecules (bottom
side elongation).} Different views of the HET-s PFD structure used as
the minimum energy structure in our simulations. Chains A and B from
the PDB fibril structure are kept. Color codes are the same as in
Fig. 2 and refer to the case of bottom side elongation.}
\end{figure*}

\begin{figure*}[!ht]
\includegraphics[]{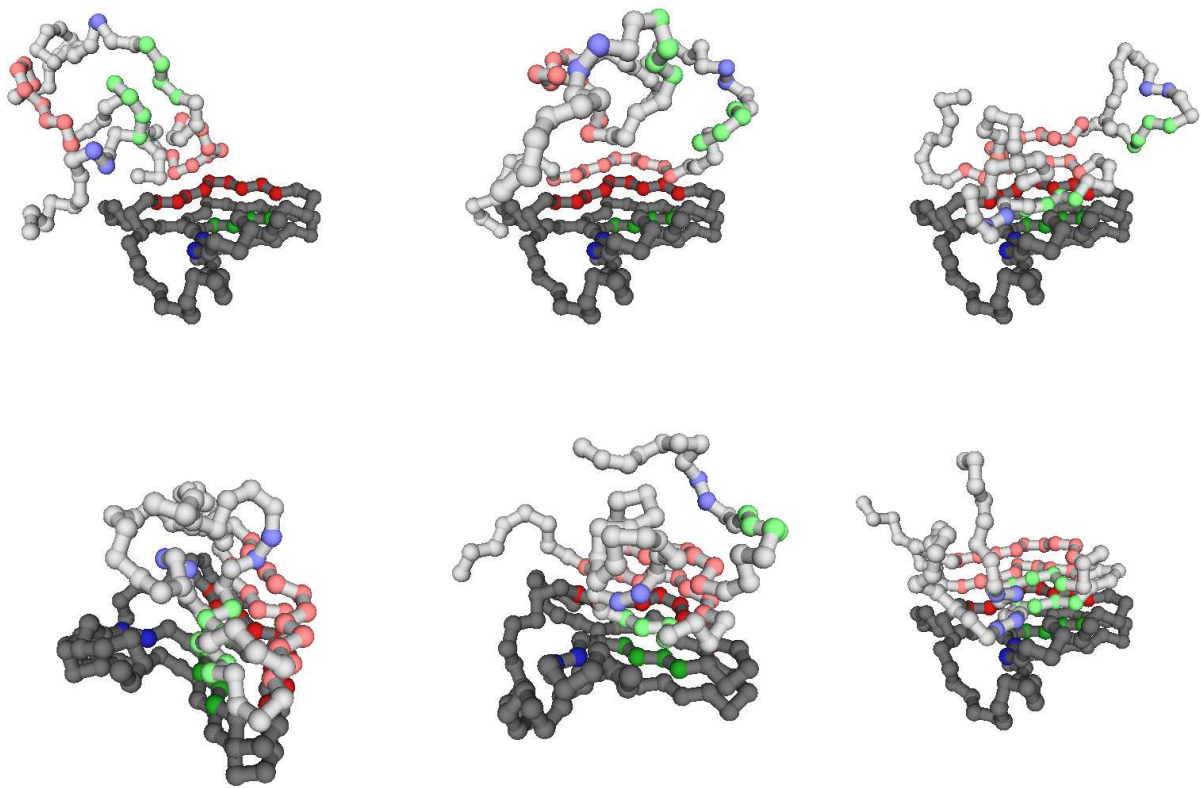}
\caption{\label{Fig4} {\bf Snapshot conformation from numerical
simulations.} Some snapshots representing conformations visited during
the simulation of fibril elongation from the top side. Color codes are
the same as in Fig. 2.}
\end{figure*}

\begin{figure*}[!ht]
\includegraphics[]{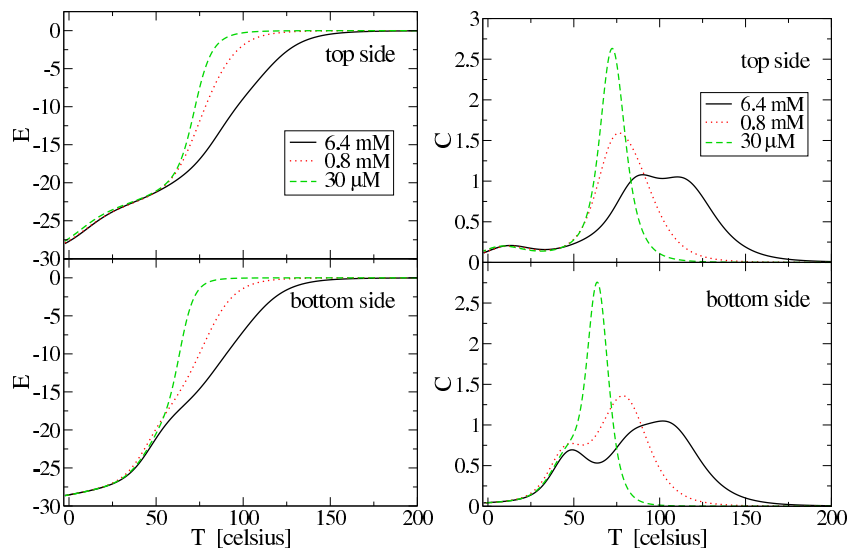}
\caption{\label{Fig5} {\bf Energy and heat capacity curves.} Energy
$E$ and heat capacity $C=dE/dT$ as a function of temperature $T$ in
$^{\circ}$C, for both elongation sides and for the three different
concentrations $c_1$, $c_2$, $c_3$ used in this
work. Both $E$ and $C$ are plotted in simulation
units so that $1$ energy unit (minus $1$ hydrogen bond) corresponds to
$3.5\ {\rm Kcal} \cdot {\rm mol}^{-1}$ (see Methods) while $1$ heat
capacity unit corresponds to the gas constant $R=1.99 \cdot 10^{-3} \
{\rm Kcal} \cdot {\rm mol}^{-1} \cdot {\rm K}^{-1}$.}
\end{figure*}

\begin{figure*}[!ht]
\includegraphics[]{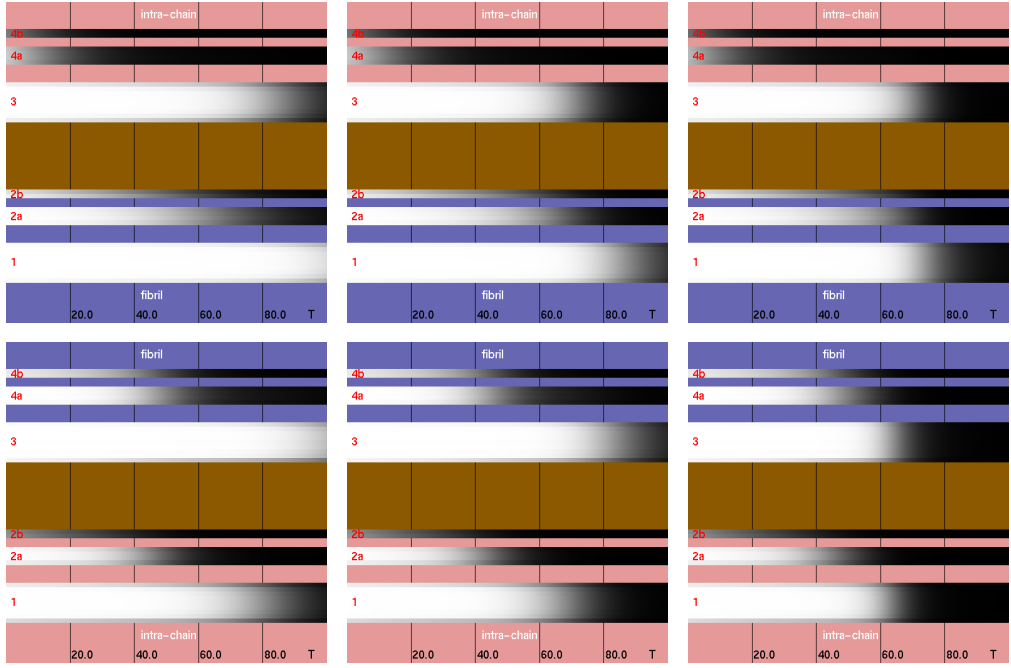}
\caption{\label{fig6} {\bf Hydrogen bond formation probabilities.}
Plots of hydrogen bond formation probabilities in the mobile chain as
a function of temperature $T$ in $^{\circ}$C (white $=$ contact formed
with probability 1, black $=$ contact never formed), for top (upper
row) and bottom (lower row) side elongation; columns refer to the
three different concentrations $c_1$ (left), $c_2$ (middle), and $c_3$
(right) used in this work. We employ a representation in which
hydrogen bonds formed between two residues are ascribed to the more
`external' residue of the two, according to the direction defined by
fibril elongation. In all plots, this residue index is represented on
the $y$-axis, from the N-terminus to the C-terminus. The lower wide
black/white band corresponds to residues in the strand $\beta_1$ (9
hydrogen bonds), followed by residues in $\beta_{2a}$ (4 hydrogen
bonds) and $\beta_{2b}$ (2 hydrogen bonds), and again residues in
$\beta_3$ (9 hydrogen bonds), followed by residues in $\beta_{4a}$ (4
hydrogen bonds) and in $\beta_{4b}$ (2 hydrogen bonds). Strands are
then divided into two larger groups, according to the nature of the
associated hydrogen bonds, which may form intra-chain (pink band) or
inter-chain (fibrillar blue band). The assignment of strands to the
two groups depends on the elongation side. For instance, for top side
elongation (upper row), strand $\beta_1$ of the mobile chain is on the
`internal' fibrillar side of the chain and the associated hydrogen
bonds form inter-chain, with the strand $\beta_3$ of the fixed
molecule. On the other hand, for bottom side elongation (lower row),
strand $\beta_1$ of the mobile chain is on the `external' exposed side
of the molecule and the associated hydrogen bonds form intra-chain,
with the strand $\beta_3$ of the same molecule. The brown band
corresponds to the loop (246-260) separating the two groups of
strands.}
\end{figure*}

\begin{figure*}[!ht]
\includegraphics[width=13cm]{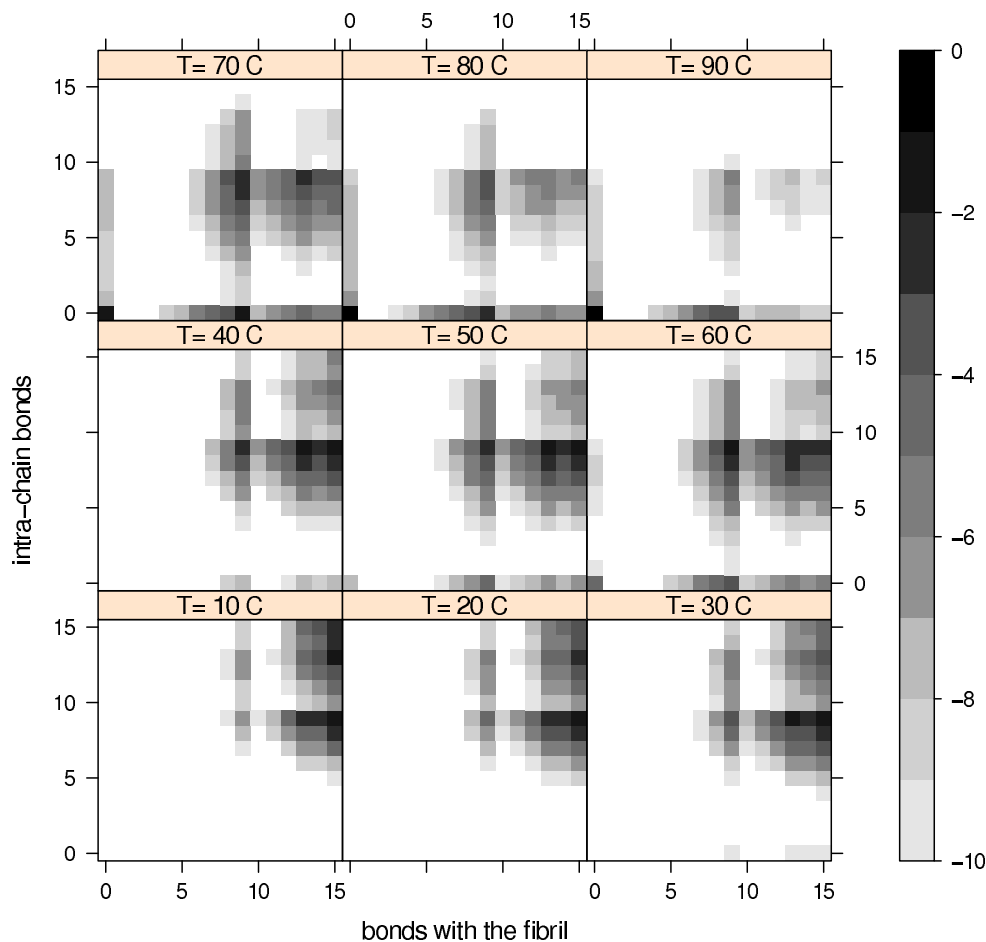}
\caption{\label{Fig7} {\bf Equilibrium occupation probabilities of
macroscopic conformational states.} Equilibrium occupation probability
of conformational states, in logarithmic scale, shown for different
temperatures $T$ in $^{\circ}$C, at concentration $c_3$ in the case of
growth from the top side. Macroscopic states are defined according to
the number of formed fibrillar and intra-chain hydrogen bonds, which
are shown in the axes of each plot. The darker (lighter) the colour,
the higher (smaller) the probability to occupy the corresponding
macroscopic state. The numbers in logarithmic scale
may be interpreted as the opposite of free energies in $RT$ units.}
\end{figure*}

\begin{figure*}[!ht]
\includegraphics[width=13cm]{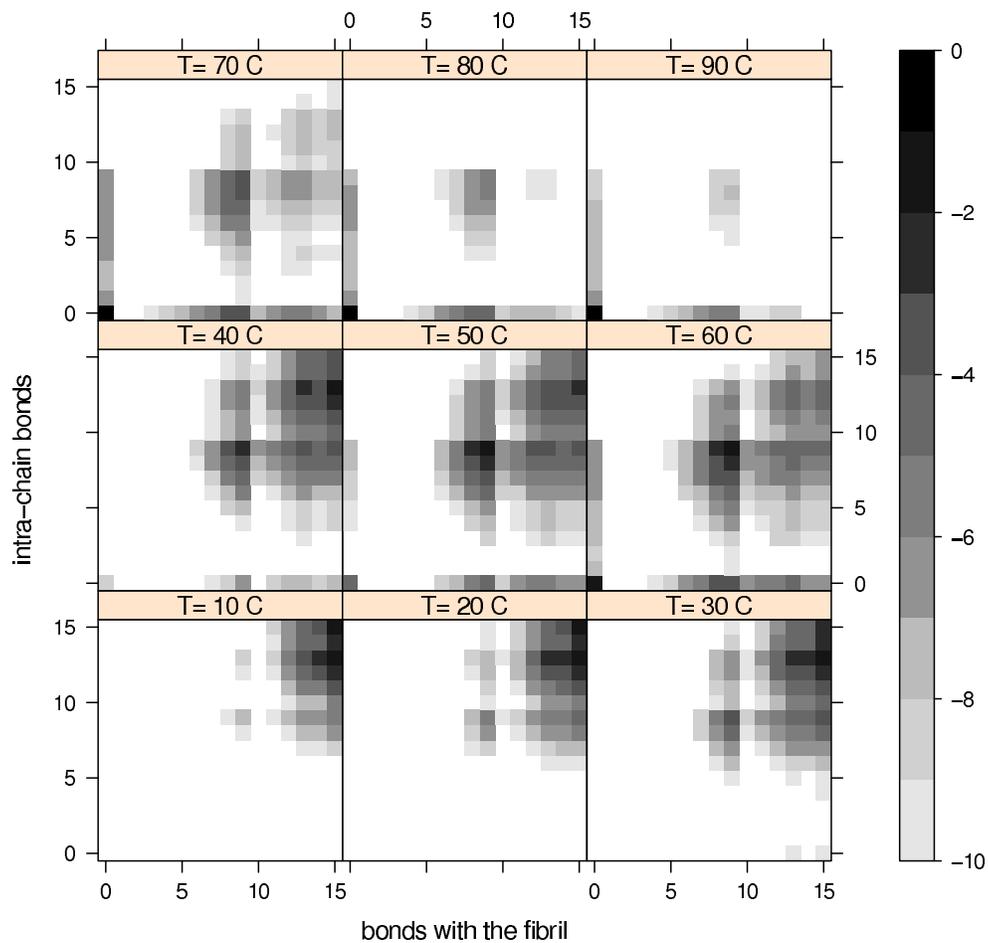}
\caption{\label{Fig8} {\bf Equilibrium occupation probabilities of macroscopic
conformational states.} As in Fig. 7, in the case of growth from the
bottom side.}
\end{figure*}

\begin{figure*}[!ht]
\includegraphics[width=13cm]{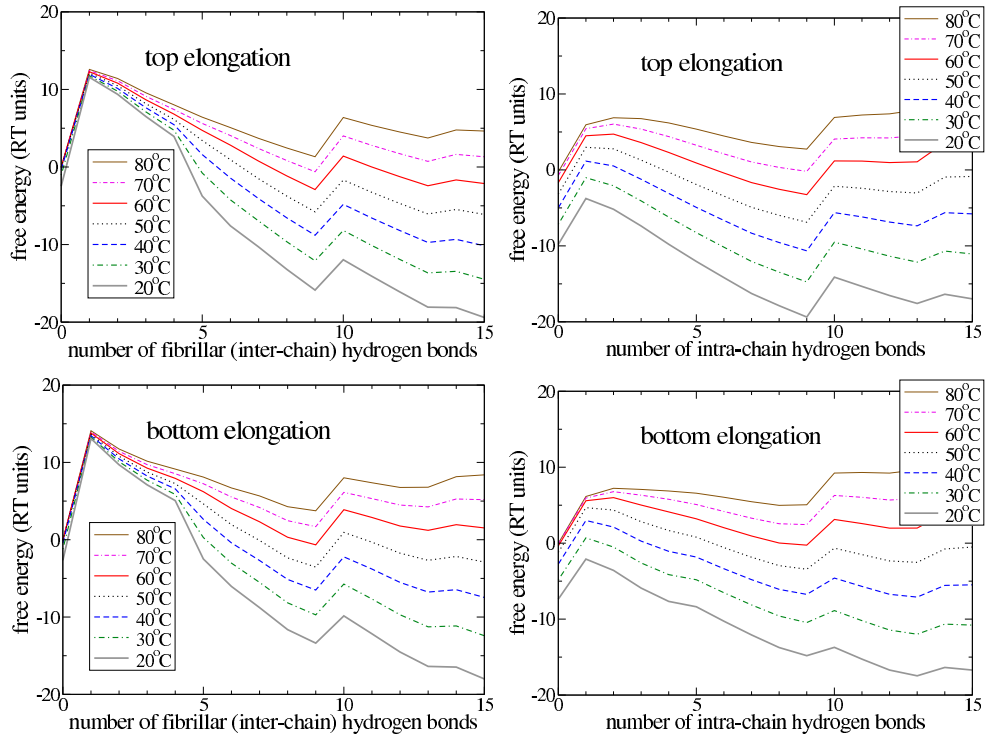}
\caption{\label{Fig9} {\bf Unidimensional energy
profiles.}  Adimensonal free energy profiles (in $RT$ units), as a
function of either the number of inter-chain fibrillar hydrogen bonds
(left panels) or the number of intra-chain hydrogen bonds (right
panels). Different profiles are shown for different temperatures in
each panel, and both elongation modes are considered for the case of
concentration $c_3$: top elongation in the upper panels and bottom
elongation in the lower panels.}
\end{figure*}

\begin{figure*}[!ht]
\includegraphics[width=13cm]{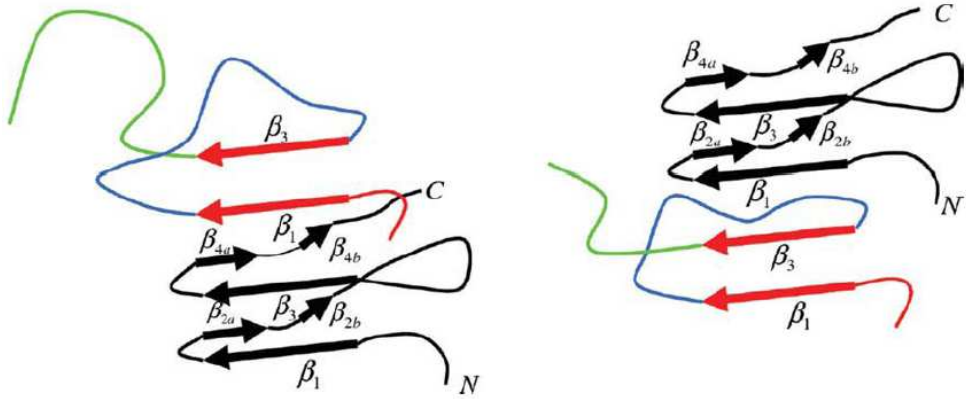}
\caption{\label{Fig10} {\bf Fibril growth polarity caused by topological
properties.} Cartoon representation of the elongating fibril after the
initial docking of the attaching mobile chain (in colours) onto the
fixed chain (in black - it represents the tip of the fibril) has just
taken place and before the final locking stage. Only $\beta$-strands
already formed are shown, in red for the mobile chain (the long
strands). The remaining fluctuating portions of the mobile chain are
shown in green (the 'tail') and in blue (the 'loop'). Right: top side
elongation. Left: bottom side elongation.}
\end{figure*}

\end{document}